\documentclass{JHEP3}

\usepackage{amsmath, amssymb}

\def\tr{\mathop{\rm tr}\nolimits}

\def\diag{\mathop{\rm diag}\nolimits}
\def\min{\mathop{\rm min}\nolimits}


\makeatletter
\def\vereq#1#2{\lower3pt\vbox{\baselineskip1.5pt \lineskip1.5pt
\ialign{$\m@th#1\hfill##\hfil$\crcr#2\crcr\sim\crcr}}}
\makeatother
\newcounter{axn}

\def\tr{\mathop{\rm tr}\nolimits}

\def\diag{\mathop{\rm diag}\nolimits}
\def\min{\mathop{\rm min}\nolimits}
\newcommand{\VEV}[1]{\left\langle #1 \right\rangle}

\newcommand{\nn}{\nonumber}

\newcommand{\eg}{{\it e.g.}}

\newcommand{\abs}[1]{\left| #1 \right|}

\newcommand{\cc}[1]{\overline{#1}}

\newcommand{\eff}{{\mrm {eff}}}
\newcommand{\symm}{{\square\!\square}}
\newcommand{\antisymm}[1]{ \begin{array}{c}
                    \square\vspace{-#1mm}\\\square
                    \end{array}}

\newcommand{\bequ}{\begin{equation}}
\newcommand{\eequ}{\end{equation}}
\newcommand{\beqn}{\begin{eqnarray}}
\newcommand{\eeqn}{\end{eqnarray}}
\newcommand{\bctr}{\begin{center}}
\newcommand{\ectr}{\end{center}}
\newcommand{\Ls}{\left(}
\newcommand{\Rs}{\right)}

\newcommand{\Ll}{\left[}
\newcommand{\Rl}{\right]}
\newcommand{\LL}{\left.}
\newcommand{\RR}{\right.}
\newcommand{\hsp}[1]{\hspace {#1cm}}
\newcommand{\vsp}[1]{\vspace {#1cm}}
\newcommand{\half}{{1\over2}}
\newcommand{\mrm}{\rm}

\def\PR#1#2#3{Phys. Rev. {\bf #1} (#3) #2 }
\def\PRL#1#2#3{Phys. Rev. Lett. {\bf #1} (#3) #2 }
\def\PL#1#2#3{Phys. Lett. {\bf #1} (#3) #2 }

\def\NP#1#2#3{Nucl. Phys. {\bf #1} (#3) #2 }

\def\PTP#1#2#3{Prog. Theor. Phys. {\bf #1} (#3)#2 }

\title{\large\bf A general formula of the effective potential 
 in 5D $SU(N)$ gauge theory on
 orbifold}

\author{
 Naoyuki Haba\\
 Institute of Theoretical Physics, 
 University of Tokushima, Tokushima 770-8502, Japan \\
}
\author{
 Toshifumi Yamashita \\
 Department of Physics, Kyoto University,
 Kyoto, 606-8502, Japan \\
}

\abstract{
We show a general formula of 
 the one loop effective potential 
 of the 5D $SU(N)$ gauge theory 
 compactified on an orbifold, $S^1/Z_2$. 
The formula shows 
 the case when 
 there are 
 fundamental, (anti-)symmetric tensor 
 and adjoint representational bulk fields. 
Our calculation method is also applicable 
 when there are bulk fields belonging to 
 higher dimensional representations. 
The supersymmetric version of 
 the effective potential 
 with Scherk-Schwarz breaking
 can be obtained straightforwardly. 
We also show some examples of 
 effective potentials 
 in $SU(3)$, $SU(5)$ and $SU(6)$
 models with various boundary
 conditions, which are 
 reproduced by our general formula.
}

\preprint{
hep-ph/0401185\\
KUNS-1896 
}

\begin{document}

\section{Introduction}

More and more people pay attention 
 to gauge 
 theories in higher dimensions. 
Especially 
 the orbifold compactification of 
 the extra dimensional spaces
 have been studied 
 by many people\cite{5d,HHHK}. 
When the gauge fields spread
 in the higher dimensions, 
 their extra dimensional components
 are regarded as 
 scalar fields
 below 
 the compactification scale. 
The zero mode of these 
 scalar fields are physical
 degrees of freedom (d.o.f.), 
 which are so-called Wilson 
 line phases. 
One of the most 
 interesting examples of using 
 Wilson line d.o.f. for the model building 
 is to regard them 
 as Higgs fields 
 in the 4D effective theory. 
This idea is so-called gauge-Higgs
 unification\cite{25,Manton:1979kb,YH,Krasnikov:dt,Lim,ghu,ghu2,ghu3,ph-gaugehiggs}. 
Here the ``adjoint Higgs fields'' can be 
 induced through the $S^1$
 compactification in 5D theory, 
 while the ``Higgs doublet fields'' 
 can be induced through the orbifold 
 compactifications. 
Another example is 
 considered in Refs.\cite{HK}, 
 where the gauge coupling unification 
 is realized due to the 
 effects of Wilson line 
 d.o.f. 
 in the non-supersymmetric (SUSY) grand unified 
 theory (GUT).

We should notice that 
 it is very
 important to calculate 
 the one loop effective potential of 
 the Wilson line d.o.f. 
 in order to determine 
 the vacuum. 
The tree level potential can not
 determine the vacuum due to the 
 existence of flat directions of 
 the Wilson line 
 phases. 
The true vacuum can be 
 determined through the 
 analysis of the effective 
 potential including the 
 quantum corrections. 
For example, 
 we can know 
 whether the color is conserved 
 or not through the analysis 
 of the one loop effective potential. 
(For the analyses 
 in 5D $SU(5)$ GUT on 
 $S^1/Z_2$, see Ref.\cite{HHHK}.) 
We can also 
 estimate the 
 finite masses of the 
 Wilson line d.o.f.
 through the 4D effective
 potential which include 
 radiative corrections. 
In general 
 the Wilson line d.o.f. 
 receive the quantum corrections, 
 and obtain the finite masses 
 of the order of the compactification scale. 
However, the quantitative estimation of 
 the finite masses is not possible until 
 we calculate the effective potential
 including the quantum corrections. 
Thus, 
 in order to determine the vacuum 
 and estimate finite masses of 
 Wilson line d.o.f.,
 the calculation of the one loop effective
 potential is strongly needed. 
(In $SU(3)$ and $SU(6)$ gauge-Higgs unification 
 models, 
 this kind of analysis 
 had been done, and 
 we found that 
 the suitable electro-weak
 symmetry breaking can be 
 realized dynamically\cite{HHKY}.)

In this paper we show a general formula of 
 the one loop effective potential 
 of the 5D $SU(N)$ gauge theory 
 compactified on an orbifold, $S^1/Z_2$. 
Although the formula only shows 
 the case when 
 there are 
 fundamental, (anti-)symmetric tensor
 and adjoint representational bulk fields, 
 our calculation method is also applicable 
 when there are bulk fields belonging to 
 higher dimensional representations. 
The SUSY version of 
 the effective potential 
 with Scherk-Schwarz (SS) breaking\cite{SS,SS3,SS4} 
 can be obtained straightforwardly. 
We also show some examples of 
 effective potentials 
 in cases of 
 one VEV with $P=P'$ (section 3), 
 two VEVs with $P=P'$ (section 4.1), 
 three VEVs with $P=P'$ (section 4.2) and 
 one VEV with $P\neq P'$ (section 4.3), 
 all of which are 
 reproduced by our general formula.

\section{The calculation method}

In the $D$ dimensional gauge theories, 
 the gauge field, $A_M$, has the 
 indices of the 4D space-time, 
 $M=$\,0\,-\,3, and extra dimensional
 directions, $M=5,6,\cdots, D$. 
The components of the 
 gauge field of extra dimensional coordinates 
 appear as scalar fields 
 in the 4D effective theory below 
 the compactification scale. 
The zero mode components of 
 these scalar fields, 
 called Wilson line phases, 
 are physical d.o.f., and 
 the estimation of the quantum corrections of them 
 is needed in order to determine 
 the vacuum and 
 finite masses. 
Here let us consider the 5D gauge theory for simplicity, 
 and show the simple calculation method 
 of evaluating the one loop effective potential 
 of the zero mode of the 5th component of 
 the 5D gauge field, $A_5$. 
Since 
 our calculation method only 
 depends on the group theoretical analysis, 
 it is available for 
 more than $5$ dimensional 
 gauge theories.

The effective potential at one loop level in a constant 
 background gauge field, $A_5$, can be obtained by 
 calculating the eigenvalues of 
 $D_M(A_5)^2=\partial_\mu^2-D_y(A_5)^2$, where 
 $D_M(A_5)$ is the covariant derivative and  
 $y$ denotes the 5th coordinate. 
The effective potential 
 induced from the gauge and ghost, 
 fermion and scalar are given by 
\begin{eqnarray}
\label{1}
&~& V_{{\eff}}[A_5]^{g+gh} = 
             -(D-2){i \over 2}\mbox{Trln} D_{M} D^{M},\\
\label{2}
&~& V_{{\eff}}[A_5]^{{\rm fermion}} = 
               f(D){i \over 2}\mbox{Trln} D_{M} D^{M},\\
&~& V_{{\eff}}[A_5]^{{\rm scalar}} = 
               - 2 {i \over 2}\mbox{Trln} D_{M} D^{M},
\label{3}
\end{eqnarray}
respectively, 
 where 
 $f(D)=2^{[D/2]}$. 
The difference of the gauge- and ghost-, fermion- and 
 scalar-contributions 
 to the effective potential  
 are only coming from 
 representations and 
 coefficients (numbers of d.o.f.) 
 in Eqs.(\ref{1})-(\ref{3}). 
For an adjoint representational field, 
 the eigenvalues are obtained 
 by diagonalizing the bilinear form, 
 ${\rm tr}\, (B  D_y(A_5)D_y(A_5) B)$, 
 where $B$ is an 
 adjoint representational field, as
\begin{equation}
-{\rm tr}\, (B  D_y(A_5)D_y(A_5) B) 
  \sim \tr (\partial_y B + ig [A_5, B])^2 . 
\label{5}
\end{equation}
Here, $[A_5, B]$ is also written as $ad(A_5)B$. 
Once the vacuum expectation value (VEV) 
 of $A_5$ is determined as 
 $\langle A_5 \rangle=aT^a$, 
 the $U(1)$ direction in the group space 
 is fixed. 
The $ad(T^a)$ is the charge operator of this $U(1)$ that 
 is generated by $T^a$. 
Thus, all we have to know is the 
 charges of this $U(1)$ 
 in order to calculate the eigenvalues 
 in Eq.(\ref{5}). 
Also, for other representations, 
 the eigenvalues of 
 $D_M(A_5)^2=\partial_\mu^2-D_y(A_5)^2$ 
 can be calculated by their $U(1)$ charges. 
We can calculate the effective potential 
 in the simple way without 
 carrying out complicated calculations of 
 the commutation relations by use of the 
 structure constants. 
In the following sections, 
 we consider the 5D $SU(N)$ gauge 
 theory compactified on 
 $S^1/Z_2$\footnote{
This calculation method is also available in the case of 
 5D coordinate being compactified on 
 $S^1$\cite{YH}.
However, the advantage of our calculation method 
 in $S^1$ case is 
 not large as that in $S^1/Z_2$ case. 
Thus, we consider only $S^1/Z_2$ case
 in this paper.}. 
In sections 3 and 4, 
 we show the effective potential 
 in some models by use of 
 our simple calculation method. 
Then in 
 section 5, 
 we show the general formula
 of the effective potential 
 of 5D $SU(N)$ gauge theory 
 compactified on $S^1/Z_2$.

\section{A simple example}

At first, we illustrate this calculation method 
 by use of a concrete simple example.
Consider a $SU(3)$ model compactified on
 $S^1/Z_2$.
We adopt boundary condition as 
\begin{equation}
 P=P'=\diag(+,-,-),
\end{equation}
which is 
 analyzed in Ref.\cite{Lim,ghu,Kaplan}. 
$P$ ($P'$) is the operator of 
 $Z_2$ transformation, $y \rightarrow -y$ 
 ($\pi R+y \rightarrow \pi R-y$). 
$R$ is the radius of the compactification 
 scale.
Under these parities, 
 $A_\mu$ and $A_5$ transform as 
\begin{eqnarray}
\label{Amu}
&&(P,P')(A_\mu)=\left(
\begin{array}{c|cc}
(+,+)&(-,-)&(-,-) \\ \cline{1-3}
(-,-)&(+,+)&(+,+) \\ 
(-,-)&(+,+)&(+,+) 
\end{array}
\right),\\
&&(P,P')(A_5) =\left(
\begin{array}{c|cc}
(-,-)&(+,+)&(+,+) \\ \cline{1-3}
(+,+)&(-,-)&(-,-) \\ 
(+,+)&(-,-)&(-,-) 
\end{array}
\right), 
\label{A5-su3}
\end{eqnarray}
which suggest 
 $SU(3)$ is broken to 
 $SU(2)\times U(1)$. 
The  Dirac fermion, $\psi$, and 
 complex scalar, $\phi$, 
 transform as
\begin{eqnarray}
\label{eta-s}
&&\phi(x, -y) = \eta T[P] \phi(x, y) ~~, ~~ 
\phi(x, \pi R-y) = \eta' T[P'] \phi(x, \pi R + y) , \\
\label{eta-D}
&&\psi(x, -y) = \eta T[P] \gamma^5 \psi(x, y) ~~, ~~
\psi(x, \pi R-y) = \eta' T[P'] \gamma^5 \psi(x, \pi R + y) ~~,
\end{eqnarray}
under the parity operators. 
$T[P]$ denotes an appropriate 
 representation matrix, for example, 
 when $\psi$ belongs to the fundamental 
 or adjoint representation, 
 $T[P]\psi$ means $P \psi$ or $P\psi P^\dagger$, 
 respectively. 
The parameters, $\eta$ and $\eta'$, are 
 like {\it intrinsic} Parity 
 eigenvalues, 
 which take $\pm 1$.

Equation (\ref{A5-su3}) suggests 
 that there is the 
 Wilson line d.o.f. as a 
 doublet of remaining 
 $SU(2)$ gauge symmetry\footnote{In SUSY case,
 there appear two doublets as 
 the Wilson line d.o.f..}. 
We can denote the VEV of 
 them as 
\begin{equation}
  \VEV{A_5}={1\over gR}
 \sum_a a_a{\lambda_a\over 2}\rightarrow \frac{1}{2gR} \Ls
    \begin{array}{ccc}
      0 & 0 & a \\
      0 & 0 & 0 \\
      a & 0 & 0 
    \end{array}\Rs,
\end{equation}
by 
 utilizing the residual global symmetry. 
Here, $g$ is the 5D gauge coupling constant.
The important point is that 
 the above VEV is proportional to one generator 
 of $SU(2)_{13}$ 
 that operates on the $2\times2$ submatrix 
 of $(1,1)$, $(1,3)$, $(3,1)$ and $(3,3)$ 
 components. 
Hereafter, we take the notation that 
 $SU(2)_{ij}$ operates 
 on the $2\times2$ submatrix 
 of $(i,i)$, $(i,j)$, $(j,i)$ and $(j,j)$ 
 components.

Now let us calculate 
 the effective potential of this $SU(3)$ model 
 induced from, for examples, 
 an adjoint and a fundamental 
 representational fields by using this 
 calculation method. 
The adjoint representation of 
 $SU(3)$ is decomposed as 
\begin{equation}
\label{ad-su3}
 \bf8 \rightarrow \bf3 +\bf1 +\bf2 +\bf2
\end{equation}
in the base of $SU(2)_{13}$.
Thus, charges of the generator are 
 given by 
\begin{equation}
\label{charge-ad-su3}
(\underbrace{+1, -1, 0}_{\bf3}, 
 \underbrace{\ 0\phantom{,} }_{\bf1}, 
 \underbrace{+1/2, -1/2}_{\bf2}, 
 \underbrace{+1/2, -1/2}_{\bf2}),
\end{equation}
since $ad(A_5)$ corresponds to 
 the $U(1)$ charge 
 defined by the $\tau_1$ direction 
 of the $SU(2)_{13}$. 
The eigenvalues of 
 $D_y(A_5)^2$ for an adjoint field $B$ become 
\begin{equation}
 2\times {n^2\over R^2}, \;\;\; 
 {(n\pm a)^2\over R^2}, \;\;\;
 2\times{(n\pm a/2)^2\over R^2},
\end{equation}
when the eigenfunctions are expanded as 
 $B \propto \cos{ny\over R}, \sin{ny\over R}$.
This Kaluza-Klein (KK)\cite{KK}
 expansion applies to 
 the gauge sector (gauge and ghost)
 and bulk fields sector 
 with $\eta\eta'=+$ in 
 Eqs.(\ref{eta-s}) and (\ref{eta-D}), 
 since 
 their parity eigenvalues, $(P,P')$, are 
 either $(+,+)$ or 
 $(-,-)$ 
 in this model. 
Then, the effective potential from 
 an adjoint representational field 
 is given by 
\begin{eqnarray}
 V_\eff^{adj(+)}&=&
 {i\over2}\int \frac{{\rm d}^4p}{(2\pi)^4}{1\over 2\pi R}
    \sum_{n=-\infty}^\infty
    \Ll \ln \Ls -p^2+\Ls\frac{n}{R}\Rs^2 \Rs
       +\ln \Ls -p^2+\Ls\frac{n-a}{R}\Rs^2 \Rs \RR \nn\\
  && \;\;\;\;\;\;\;\;\;\;\;\;\;\;\;\;\;\;\;\;\;\;\;\;
    \LL  +2\ln \Ls -p^2+\Ls\frac{n-a/2}{R}\Rs^2 \Rs
    \Rl  \nn \\
  &=&
    {C\over2} \sum_{n=1}^\infty \frac{1}{n^5} \;
     [\cos(2\pi na)+2\cos(\pi na)],
\label{V-su3}
\end{eqnarray}
where $C\equiv 3/(64\pi^7R^5)$. 
The 2nd equation is derived by the 
 Wick rotation and omitting 
 independent terms of the VEV, $a$\cite{HHHK,Lim}. 
Equation (\ref{V-su3}) shows the 
 effective potential from 
 one (fermionic) d.o.f. of the field, 
 so that 
 the true 
 effective potential 
 is obtained by producting 
 coefficients as, 
 [(fermionic d.o.f.) $-$ (bosonic d.o.f.)] $\times$ 
 Eq.(\ref{V-su3}).
For examples, 
 the 
 gauge sector 
 contribution is 
 $-3 \times$ Eq.(\ref{V-su3}), 
 which correctly 
 reproduces the result in Refs.\cite{Lim,HHKY}.  
As for the contributions from 
 bulk fields sector, 
 a Dirac fermion and a complex scalar 
 with $\eta\eta'=+$, 
 are obtained by 
 producting coefficients, $4$ and $-2$, 
 in Eq.(\ref{V-su3}), respectively.

On the other hand, 
 when 
 the adjoint representational 
 bulk fields have $\eta \eta'=-$ 
 in Eqs.(\ref{eta-s}) and (\ref{eta-D}), 
 the eigenfunctions are expanded as 
 $B\propto\cos{(n+1/2)y\over R},\sin{(n+1/2)y\over R}$  
 because their parity eigenvalues are 
 $(P,P')=(+,-)$ or $(-,+)$. 
These half KK 
 expansions  induce 
 a sift of VEV $\VEV{A_5}$ as
\begin{equation}
 \Ls\frac{\Ls n+\half\Rs+Qa}{R}\Rs = 
           \Ls\frac{n+\Ls Qa+\half\Rs}{R}\Rs,
\label{Q}
\end{equation}
where $Q$ is the 
 $U(1)$ charge of 
 definite representation of $B$. 
Then, in the case of 
 the adjoint representation with $\eta \eta'=-$, 
 the eigenvalues of 
 $D_y(A_5)^2$ become 
\begin{equation}
 2\times {(n+1/2)^2\over R^2}, \;\;\; 
 {(n\pm a+1/2)^2\over R^2}, \;\;\;
 2\times{(n\pm a/2+1/2)^2\over R^2},
\end{equation}
which induce 
\begin{eqnarray}
 V_\eff^{adj(-)}
  &=&{i\over2}\int \frac{{\rm d}^4p}{(2\pi)^4}{1\over 2\pi R}
    \sum_{n=-\infty}^\infty
    \Ll \ln \Ls -p^2+\Ls\frac{n+1/2}{R}\Rs^2 \Rs \RR \nn\\
  &&     +\ln \Ls -p^2+\Ls\frac{n-a+1/2}{R}\Rs^2 \Rs 
    \LL  +2\ln \Ls -p^2+\Ls\frac{n-a/2+1/2}{R}\Rs^2 \Rs
    \Rl  \nn \\
  &=&
    {C\over2} \sum_{n=1}^\infty \frac{1}{n^5}\;
     [\cos(2\pi n(a-1))+2\cos(\pi n(a-1))].
\label{V-su3-}
\end{eqnarray}

For other representations, 
 the eigenvalues of the covariant derivative can be also 
 evaluated from charges of the same $U(1)$. 
Let us consider the contribution from a 
 fundamental representational field, for an example. 
The fundamental representation is 
 decomposed 
 as 
\begin{equation}
\label{fund-su3}
 \bf3 \rightarrow \bf2 +\bf1 
\end{equation}
in the base of $SU(2)_{13}$. 
Then, the $U(1)$ charges 
 corresponding to 
 Eq.(\ref{charge-ad-su3}) are 
\begin{equation}
\label{charge-f-su3}
(\underbrace{+1/2, -1/2}_{\bf2}, 
 \underbrace{\phantom{,}0}_{\bf1}).
\end{equation}
This means that 
 a fundamental field $B$ with $\eta\eta'=+$ 
 has eigenvalues of 
 $D_y(A_5)^2$ as 
\begin{equation}
{(n\pm a)^2\over R^2}, \;\;\; {n^2\over R^2}\;. 
\end{equation}
Here eigenfunctions are expanded as  
 $B\propto \cos{ny\over R}, \sin{ny\over R}$, 
 since $(P,P')=(+,+)$ or $(-,-)$.  
Thus, the effective potential from 
 the fundamental representational field 
 is given by
\begin{eqnarray}
  V_\eff^{fnd(+)}
  &=&{C\over2} \sum_{n=1}^\infty \frac{1}{n^5}\;
     \cos(\pi na).  
\label{V-su3-f}
\end{eqnarray}
In case of $\eta \eta'=-$,  
 eigenfunctions are expanded as 
 $B\propto\cos{(n+1/2)y\over R}, \sin{(n+1/2)y\over R}$,
 since $(P,P')=(+,-)$ or $(-,+)$.  
This suggests 
 the eigenvalues of 
 $D_y(A_5)^2$ become 
\begin{equation}
{(n\pm a+1/2)^2\over R}, \;\;\; {(n+1/2)^2\over R}, 
\end{equation}
which 
 induces the effective potential, 
\begin{eqnarray}
 V_\eff^{fnd(-)}
\label{V-su3-f-}
  &=&{C\over2} \sum_{n=1}^\infty \frac{1}{n^5}\;
     \cos(\pi n(a-1)). 
\end{eqnarray}

Now let us show the case 
 when there are $N_a^{(\pm)}$ ($N_f^{(\pm)}$) numbers
 of adjoint (fundamental) Dirac fermions and 
 $N_s^{(\pm)}$ numbers 
 of complex scalars 
 of fundamental representation
 in the bulk. 
The index $(\pm)$ denotes the 
 eigenvalue of $\eta\eta'$. 
The discussion below 
 Eq.(\ref{V-su3}) suggests 
 the gauge sector induces the 
 effective potential, 
 $-3 \times$ Eq.(\ref{V-su3}). 
While bulk fields contributions 
 are $4N_a^{(+)} \times$ Eq.(\ref{V-su3}), 
 $4N_a^{(-)} \times$ Eq.(\ref{V-su3-}), 
 $(4N_f^{(+)}-2N_s^{(+)}) \times$ Eq.(\ref{V-su3-f}), and 
 $(4N_f^{(-)}-2N_s^{(-)}) \times$ Eq.(\ref{V-su3-f-}). 
It is because 
 Eqs.(\ref{V-su3}), (\ref{V-su3-}), (\ref{V-su3-f}), and 
 (\ref{V-su3-f-}) show the 
 effective potential from 
 one (fermionic) d.o.f. of the field, 
 so that 
 the true 
 effective potential 
 is obtained by producting 
 coefficients, 
 [(fermionic d.o.f.) $-$ (bosonic d.o.f.)]. 
Thus, 
 the total effective potential 
 becomes 
\begin{eqnarray}
\label{Vsu3-total}
V_{\eff}
&=& C\sum_{n=1}^{\infty}{1\over n^5}
    [(-{3\over2}+2 N_a^{(+)}) \cos (2\pi na) + 
     2 N_a^{(-)} \cos (\pi n(2a-1)) \nonumber \\
&+& (-3+4 N_a^{(+)}-N_s^{(+)} + 2 N_f^{(+)}) 
    \cos(\pi na) \nonumber \\
&+& (4 N_a^{(-)}-N_s^{(-)} + 2 N_f^{(-)}) \cos(\pi n(a-1))], 
\end{eqnarray}
which reproduces the results 
 in Ref.\cite{HHKY}. 
This counting rule for the 
 coefficients of the 
 effective potential 
 is applicable in general. 
Thus, we show the 
 effective potential induced from 
 one d.o.f. of each representational field 
 in the following discussions.

The SUSY version of the 
 effective potential can be obtained 
 straightforwardly,
 when 
 SUSY breaking is induced 
 by the SS 
 mechanism\footnote{In the case of 
 other SUSY breaking, such as 
 introducing explicit soft breaking 
 masses\cite{Takenaga:2003dp}, 
 the calculation of the effective potential 
 might be easily done 
 in a similar manner.}. 
{}For the gauge and ghost contributions, 
 the coefficient 
 should be modified as 
 $-3 \times$ Eq.(\ref{V-su3})
 $\rightarrow$
 $-4 \times$ Eq.(\ref{V-su3}), 
 and 
 the factor $(1-\cos(2\pi n\beta))$ 
 should be added in the r.h.s. summation 
 in Eq.(\ref{V-su3}). 
They are coming from 
 {\it massive} gaugino contributions.   
The $\beta$ parameterizes SS SUSY breaking, 
 and 4D effective theory has the gaugino mass 
 of 
 order 
 $\beta /R$\cite{HHHK}. 
As for the bulk fields, 
 they are corresponding to 
 the hypermultiplets of 4D ${\cal N}=2$ SUSY. 
Since 
 one hypermultiplet has 
 one Dirac fermion and two complex
 scalar 
 d.o.f., 
 and 
 scalar components always have
 SUSY breaking masses, 
 the SUSY effective potential 
 is obtained by 
 adding the factor $(1-\cos(2\pi n\beta ))$ 
 in the summation $n$ of 
 the non-SUSY effective potential induced from 
 the Dirac fermion 
 contributions. 
By using this 
 technique, we can obtain 
 the effective potential in the 
 SUSY version of 
 this $SU(3)$ model. 
Considering the situation that 
 $N_f^{(\pm)}$ and $N_a^{(\pm)}$ 
 species of hypermultiplets of 
 fundamental and 
 adjoint representations in 
 the bulk, respectively, 
 the effective potential becomes 
\begin{eqnarray}
V_{\eff}
&=& 2 C\sum_{n=1}^{\infty}{1\over n^5}
         (1-\cos(2\pi n\beta)) \times 
 [(N_a^{(+)}-1)\cos(2\pi na)+N_a^{(-)}
  \cos(\pi n(2a-1))\nonumber \\
&& +   (2N_a^{(+)} +N_f^{(+)} -2) \cos(\pi na)+
       (2N_a^{(-)} +N_f^{(-)}) \cos(\pi n(a-1))], 
\label{Vsu3-SUSY}
\end{eqnarray}
which reproduces 
 the result in Ref.\cite{HHKY}.

\section{More general examples}

In this section we show more complicated 
 examples. 
We calculate the effective potential 
 when there are non-vanishing 
 two or three VEVs in $\langle A_5 \rangle$, 
 and also the case of $P \neq P'$.

\subsection{Two VEVs with $P=P'$}

Here 
 we show an example of 
 existing two VEVs with $P=P'$ 
 in the $SU(5)$ GUT model\cite{HHHK}.  
This model has parities, 
\begin{eqnarray}
  P=P'=\diag(1,1,1,-1,-1), 
\end{eqnarray}
under which 
 4D gauge field 
 transforms as 
\begin{equation}
(P,P')(A_\mu)=\left(
\begin{array}{ccc|cc}
(+,+)&(+,+)&(+,+)&(-,-)&(-,-)\\
(+,+)&(+,+)&(+,+)&(-,-)&(-,-)\\
(+,+)&(+,+)&(+,+)&(-,-)&(-,-)\\ \cline{1-5}
(-,-)&(-,-)&(-,-)&(+,+)&(+,+)\\
(-,-)&(-,-)&(-,-)&(+,+)&(+,+)
\end{array}
\right). 
\label{su5-parity}
\end{equation}
This means that 
 the gauge symmetry is reduced as 
 $SU(5)\rightarrow 
 SU(3)_c\times SU(2)_L\times U(1)_Y$. 
Since the signs of parities in 
 each component of $A_5$ 
 are completely opposite 
 to those of $A_\mu$
 as shown in 
 Eqs.(\ref{Amu}) and (\ref{A5-su3}),  
 the zero modes 
 exist in upper-right $3\times2$ and 
 lower-left $2\times3$ submatrices 
 corresponding to $(-,-)$ in
 Eq.(\ref{su5-parity}). 
By using the residual global 
 symmetry, the d.o.f. of 
 Wilson line phases can be set two
 as 
\begin{equation}
\label{su5-2}
  \VEV{A_5} = \frac{1}{2gR} \Ls
    \begin{array}{ccccc}
      0 & 0 & 0 & 0 & a \\
      0 & 0 & 0 & b & 0 \\
      0 & 0 & 0 & 0 & 0 \\
      0 & b & 0 & 0 & 0 \\
      a & 0 & 0 & 0 & 0 \\
    \end{array}\Rs.
\end{equation}

In order to obtain the effective potential, 
 it is important to know the eigenvalues of 
 $D_y(A_5)^2$ for 
 the following two generators:
\[
   \half\Ls
    \begin{array}{ccccc}
      0 & 0 & 0 & 0 & 1 \\
      0 & 0 & 0 & 0 & 0 \\
      0 & 0 & 0 & 0 & 0 \\
      0 & 0 & 0 & 0 & 0 \\
      1 & 0 & 0 & 0 & 0 \\
    \end{array}\Rs, \quad
   \half\Ls
    \begin{array}{ccccc}
      0 & 0 & 0 & 0 & 0 \\
      0 & 0 & 0 & 1 & 0 \\
      0 & 0 & 0 & 0 & 0 \\
      0 & 1 & 0 & 0 & 0 \\
      0 & 0 & 0 & 0 & 0 \\
    \end{array}\Rs.
\]
They are generators ($\tau_1$) of 
 $SU(2)_{15}$ and $SU(2)_{24}$, respectively.
As in the section 3, 
 let us see the decomposition of 
 $SU(5)$ into
 $SU(2)_{15} \times SU(2)_{24}$.

The adjoint representation of 
 $SU(5)$ is decomposed as 
\begin{equation}
  {\bf24} \rightarrow 
     ({\bf3},{\bf1})+({\bf1},{\bf3})+2\times({\bf1},{\bf1})
    +2\times({\bf2},{\bf1})+2\times({\bf1},{\bf2})
    +2\times({\bf2},{\bf2}).
\end{equation}
This means the eigenvalues of 
 $D_y(A_5)^2$ for the adjoint field are 
\begin{equation}
4\times {n^2\over R^2},
 \quad {(n\pm a)^2\over R^2},\quad {(n\pm b)^2\over R^2},\quad 
 2\times{(n\pm a/2)^2\over R^2}, 
 \quad 2\times{(n\pm b/2)^2\over R^2}, 
 \quad 2\times{(n\pm (a \pm b)/2)^2\over R^2},
\end{equation}
for $\cos{ny\over R}$ and $\sin{ny\over R}$
 modes. 
They can be also obtained by 
 the $U(1)$ charges, which are calculated 
 by the commutation relations in the gauge 
 where $\langle A_5 \rangle$ is diagonal, 
 $\VEV{A_5}\propto\diag(a,b,0,-b,-a)$.
In this gauge, charge of each component is given by 
\[
  Q(A_\mu) = \Ls
    \begin{array}{ccccc}
      0 & (a-b)/2 & a/2 & (a+b)/2 & a \\
      (-a+b)/2 & 0 & b/2 & b & (a+b)/2 \\
      -a/2 & -b/2 & 0 & b/2 & a/2 \\
      (-a-b)/2 & -b & -b/2 & 0 & (a-b)/2 \\
      -a & (-a-b)/2 & -a/2 & (-a+b)/2 & 0 \\
    \end{array}\Rs.
\]
These eigenvalues suggest that 
 the contribution from 
 one d.o.f. 
 of the adjoint representational field 
 becomes 
\beqn
  V_\eff^{{adj(+)}} &=& 
     \frac{C}{2}\sum_{n=1}^\infty \frac{1}{n^5} 
     \Ll \cos(2\pi na)+\cos(2\pi nb)+
     2\cos(\pi na)+2\cos(\pi nb) \RR\nn\\
   &&\LL\hsp{1.5}+2\cos(\pi n(a+b))+                             
                  2\cos(\pi n(a-b)) \Rl .
\label{V-su5}
\eeqn

As for $\overline{\bf 5}$ and 
 ${\bf 10}$ representations 
 with $\eta\eta'=+$, 
 the decompositions in terms of 
 $SU(2)_{15} \times SU(2)_{24}$ 
 are given as 
\beqn
  \overline{\bf 5} &\rightarrow& ({\bf1},{\bf1})+({\bf2},{\bf1})
                      +({\bf1},{\bf2}), 
\eeqn
and
\beqn
  {\bf 10}&\rightarrow& 2\times({\bf1},{\bf1})
                      +({\bf2},{\bf1})+({\bf1},{\bf2})
                      +({\bf2},{\bf2}).
\eeqn
They mean that 
 eigenvalues of $D_y(A_5)^2$ are 
\beqn
&& {n^2\over R^2},
   \quad {(n\pm a/2)^2\over R^2},
   \quad {(n\pm b/2)^2\over R^2}, 
\eeqn
and 
\beqn
&& 2\times{n^2\over R^2}, \quad 
   \quad {(n\pm a/2)^2\over R^2},\quad 
         {(n\pm b/2)^2\over R^2}, 
   \quad {(n\pm (a \pm b)/2)^2\over R^2},
\eeqn
respectively. 
Thus, the contributions from 
 one d.o.f. of 
 $\overline{\bf 5}$ and ${\bf 10}$ 
 with $\eta\eta'=+$
 are given by 
\beqn
\label{su5-5}
  V_\eff^{\overline{\bf 5}(+)} &=& 
     \frac{C}{2}\sum_{n=1}^\infty \frac{1}{n^5} 
     \Ll \cos(\pi na)+\cos(\pi nb) \Rl , \\
\label{su5-10}
  V_\eff^{{\bf 10}(+)} &=& 
     \frac{C}{2}\sum_{n=1}^\infty \frac{1}{n^5} 
     \Ll \cos(\pi na)+\cos(\pi nb) +
     \cos(\pi n(a+b))+\cos(\pi n(a-b)) \Rl,
\eeqn
 respectively, 
 which also reproduce the result in 
 Ref.\cite{HHHK} correctly.

Notice that 
 the contribution from one 
 d.o.f. 
 with $\eta\eta'=-$ 
 is easily obtained by 
 $Qa+Qb \rightarrow Qa+Qb+\half$
 in Eqs.(\ref{V-su5}), (\ref{su5-5}),
 and (\ref{su5-10}) 
 as discussed in section 3. 
The results are given as
\beqn
  V_\eff^{{adj(-)}} &=& 
     \frac{C}{2}\sum_{n=1}^\infty \frac{1}{n^5} 
     \Ll \cos(\pi n(2a-1))+\cos(\pi n(2b-1)) \RR\nn\\
   &&\LL\hsp{1.5}+2\cos(\pi n(a-1))+2\cos(\pi n(b-1)) \RR\nn\\
   &&\LL\hsp{1.5}+2\cos(\pi n(a+b-1))+                             
                  2\cos(\pi n(a-b-1)) \Rl ,  \\
  V_\eff^{\overline{\bf 5}(-)} &=& 
     \frac{C}{2}\sum_{n=1}^\infty \frac{1}{n^5} 
     \Ll \cos(\pi n(a-1))+\cos(\pi n(b-1)) \Rl , \\
  V_\eff^{{\bf 10}(-)} &=& 
     \frac{C}{2}\sum_{n=1}^\infty \frac{1}{n^5} 
     \Ll \cos(\pi n(a-1))+\cos(\pi n(b-1)) \RR\nn\\
     &&\LL\hsp{1.5}+\cos(\pi n(a+b-1))+\cos(\pi n(a-b-1)) \Rl .
\eeqn

\subsection{Three VEVs with $P=P'$}

Next, 
 we show the example of 
 existing three VEVs with $P=P'$ 
 in the $SU(6)$ GUT model. 
This model has the parities, 
\begin{eqnarray}
  P=P'=\diag(1,1,1,-1,-1,-1), 
\end{eqnarray}
under which 
 4D gauge field 
 transforms 
\begin{equation}
(P,P')(A_\mu)=\left(
\begin{array}{ccc|ccc}
(+,+)&(+,+)&(+,+)&(-,-)&(-,-)&(-,-)\\
(+,+)&(+,+)&(+,+)&(-,-)&(-,-)&(-,-)\\
(+,+)&(+,+)&(+,+)&(-,-)&(-,-)&(-,-)\\ \cline{1-6}
(-,-)&(-,-)&(-,-)&(+,+)&(+,+)&(+,+)\\
(-,-)&(-,-)&(-,-)&(+,+)&(+,+)&(+,+)\\
(-,-)&(-,-)&(-,-)&(+,+)&(+,+)&(+,+)
\end{array}
\right). 
\label{su6-parity}
\end{equation}
This means that 
 the gauge symmetry is reduced as 
 $SU(6)\rightarrow SU(3)_c\times SU(3)_L\times U(1)$. 
Since the signs of parities in 
 each component of $A_5$ 
 are completely opposite to 
 those of $A_\mu$,  
 the zero modes 
 exist in upper-right $3\times3$ and 
 lower-left $3\times3$ submatrices
 corresponding to $(-,-)$ in
 Eq.(\ref{su6-parity}). 
By using the residual global 
 symmetry, the d.o.f. of 
 the Wilson line phases can be set three 
 as 
\begin{equation}
\label{su6-1}
  \VEV{A_5} = \frac{1}{2gR} \Ls
    \begin{array}{cccccc}
      0 & 0 & 0 & 0 & 0 & a \\
      0 & 0 & 0 & 0 & b & 0 \\
      0 & 0 & 0 & c & 0 & 0 \\
      0 & 0 & c & 0 & 0 & 0 \\
      0 & b & 0 & 0 & 0 & 0 \\
      a & 0 & 0 & 0 & 0 & 0 \\
    \end{array}\Rs.
\end{equation}

The effective potential can 
 be calculated in a similar
 manner as in the section 4.1. 
The adjoint representation of 
 $SU(6)$ is decomposed as 
\beqn
  {\bf35} &\rightarrow& 
      ({\bf3},{\bf1},{\bf1})+({\bf1},{\bf3},{\bf1})+({\bf1},{\bf1},{\bf3})
     +2\times({\bf1},{\bf1},{\bf1}) \nn\\
  && +2\times({\bf2},{\bf2},{\bf1})+2\times({\bf1},{\bf2},{\bf2})
      +2\times({\bf2},{\bf1},{\bf2}), 
\eeqn
in terms of 
 $SU(2)_{16} \times SU(2)_{25} \times SU(2)_{34}$.
This leads the contribution from 
 one d.o.f. of 
 the adjoint representational field 
 as 
\beqn
  V_\eff^{{adj(+)}} &=& 
     \frac{C}{2}\sum_{n=1}^\infty \frac{1}{n^5} 
     \Ll \cos(2\pi na)+\cos(2\pi nb)+
         \cos(2\pi nc) \RR\nn\\
   &&\LL\hsp{1.5}     
 +2\cos(\pi n(a+b))+2\cos(\pi n(a-b)) \RR\nn\\
   &&\LL\hsp{1.5}     
 +2\cos(\pi n(b+c))+2\cos(\pi n(b-c)) \RR\nn\\
   &&\LL\hsp{1.5}     
 +2\cos(\pi n(c+a))+2\cos(\pi n(c-a)) \Rl .
\eeqn

As for a fundamental representational field 
 with 
 $\eta \eta'=+$, 
 the decomposition is given as following:
\[
 {\bf6} \rightarrow 
    ({\bf2},{\bf1},{\bf1})+({\bf1},{\bf2},
     {\bf1})+({\bf1},{\bf1},{\bf2})
\]
This means 
 the contribution from 
 one d.o.f of 
 the fundamental representational 
 field with $\eta\eta'=+$ 
 is 
\bequ
  V_\eff^{{fnd(+)}} = 
     \frac{C}{2}\sum_{n=1}^\infty \frac{1}{n^5} 
     \Ll \cos(\pi na)+\cos(\pi nb)+
         \cos(\pi nc) \Rl .
\eequ
As for 
 the contribution from fields 
 with $\eta\eta'=-$, 
 the effective potential is obtained by 
 $Qa+Qb+Qc \rightarrow Qa+Qb+Qc+\half$
 as discussed in section 3. 
They are shown as 
\beqn
  V_\eff^{{adj(-)}} &=& 
     \frac{C}{2}\sum_{n=1}^\infty \frac{1}{n^5} 
     \Ll \cos(\pi n(2a-1))+\cos(\pi n(2b-1))+
         \cos(\pi n(2c-1)) \RR\nn\\
   &&\LL\hsp{1.5}     
 +2\cos(\pi n(a+b-1))+2\cos(\pi n(a-b-1)) \RR\nn\\
   &&\LL\hsp{1.5}     
 +2\cos(\pi n(b+c-1))+2\cos(\pi n(b-c-1)) \RR\nn\\
   &&\LL\hsp{1.5}     
 +2\cos(\pi n(c+a-1))+2\cos(\pi n(c-a-1)) \Rl , \\
  V_\eff^{{fnd(-)}} &=& 
     \frac{C}{2}\sum_{n=1}^\infty \frac{1}{n^5} 
     \Ll \cos(\pi n(a-1))+\cos(\pi n(b-1))+
         \cos(\pi n(c-1)) \Rl .
\eeqn

In section 4.1 and 4.2, 
 we have shown the $P=P'$ case, 
 where only either 
 $(+,+)$ and $(-,-)$ modes 
 (in gauge sector and bulk fields sector with 
 $\eta\eta'=+$), or
 $(+,-)$ and $(-,+)$ modes 
 (in bulk fields sector with 
 $\eta\eta'=-$) 
 exist in each 
 representation. 
In section 4.3, 
 we will show an example 
 of $P\neq P'$ case, where
 all modes of $(\pm,\pm)$ 
 can exist in one representation.

\subsection{One VEV with $P\neq P'$}

Now let us show 
 an example of 
 existing one VEV with $P\neq P'$ 
 in the $SU(6)$ GUT model\cite{ghu,HHKY}.  
This model has the parities, 
\begin{eqnarray}
\label{PP'-su6}
  P&=&\diag(1,1,1,1,-1,-1) \nn\\
  P'&=&\diag(1,-1,-1,-1,-1,-1),
\end{eqnarray}
under which 
 4D gauge field 
 transforms 
\begin{equation}
(P,P')(A_\mu)=\left(
\begin{array}{c|ccc|cc}
(+,+)&(+,-)&(+,-)&(+,-)&(-,-)&(-,-)\\ \cline{1-6}
(+,-)&(+,+)&(+,+)&(+,+)&(-,+)&(-,+)\\ 
(+,-)&(+,+)&(+,+)&(+,+)&(-,+)&(-,+)\\
(+,-)&(+,+)&(+,+)&(+,+)&(-,+)&(-,+)\\ \cline{1-6}
(-,-)&(-,+)&(-,+)&(-,+)&(+,+)&(+,+)\\
(-,-)&(-,+)&(-,+)&(-,+)&(+,+)&(+,+)
\end{array}
\right). 
\label{su6-parity-2}
\end{equation}
This means that 
 the gauge symmetry is reduced as 
 $SU(6)\rightarrow SU(3)_c\times 
 SU(2)_L\times U(1)_Y\times U(1)$. 
Since the signs of parities in 
 each component of $A_5$ 
 are completely opposite to those of $A_\mu$, 
 the zero modes 
 exist in upper-right $1\times2$ and 
 lower-left $2\times1$ submatrices
 corresponding to $(-,-)$ in
 Eq.(\ref{su6-parity-2}). 
This zero mode is regarded
 as a ``Higgs doublet'' 
 in the gauge-Higgs 
 unified models\cite{ghu,HHKY}. 
By using the residual global 
 symmetry, the d.o.f. of 
 the Wilson line phase can be set just 
 one as 
\begin{equation}
\label{su6-2}
  \VEV{A_5} = \frac{1}{2gR} \Ls
    \begin{array}{cccccc}
      0 & 0 & 0 & 0 & 0 & a \\
      0 & 0 & 0 & 0 & 0 & 0 \\
      0 & 0 & 0 & 0 & 0 & 0 \\
      0 & 0 & 0 & 0 & 0 & 0 \\
      0 & 0 & 0 & 0 & 0 & 0 \\
      a & 0 & 0 & 0 & 0 & 0 \\
    \end{array}\Rs.
\end{equation}

We can always take the gauge, 
 in which this VEV becomes 
 diagonal as
 $\VEV{A_5} \propto$ $\diag(1,0,0,0,0,-1)$. 
The $U(1)$ charge, $Q$, for this direction
 and the eigenvalue of $PP'$ 
 are given by
\begin{equation}
(Q,PP')(A_\mu)=\left(
\begin{array}{cccccc}
(0,+)&(\half,-)&(\half,-)&(\half,-)&(\half,+)&(1,+)\\ 
(-\half,-)&(0,+)&(0,+)&(0,+)&(0,-)&(\half,-)\\ 
(-\half,-)&(0,+)&(0,+)&(0,+)&(0,-)&(\half,-)\\
(-\half,-)&(0,+)&(0,+)&(0,+)&(0,-)&(\half,-)\\ 
(-\half,+)&(0,-)&(0,-)&(0,-)&(0,+)&(\half,+)\\
(-1,+)&(-\half,-)&(-\half,-)&(-\half,-)&(-\half,+)&(0,+)
\end{array}
\right).
\end{equation}
This means that the eigenvalues of 
 $D_y(A_5)^2$ are 
\begin{equation}
11\times {n^2\over R^2}, \;\; 
 6\times {(n+1/2)^2\over R^2}, \;\; 
{(n\pm a)^2\over R}, \;\;\;
 2\times{(n\pm a/2)^2\over R^2}, \;\; 
 6\times{(n\pm a/2+1/2)^2\over R^2}, 
 \label{53}
\end{equation}
since  
 the eigenfunctions are expanded as 
 $B \propto \cos{ny\over R}, \sin{ny\over R}$
 ($\cos{(n+1/2)y\over R}, \sin{(n+1/2)y\over R}$)
 for $PP'=+$ ($PP'=-$). 
This case is applicable 
 for the gauge sector and bulk fields 
 sector with $\eta\eta'=+$. 
Equation (\ref{53}) suggests 
 that the effective potential for
 one d.o.f. is given by 
\begin{equation}
 V_\eff^{adj(+)} ={C\over2} \sum_{n=1}^\infty \frac{1}{n^5} \Ll
     6\cos(n\pi(a-1))+2\cos(n\pi a)+\cos(2n\pi a) \Rl .
\end{equation}
We can also reach this conclusion via an analysis 
 independent of matrix representation. 
The product of parities (Eq.(\ref{PP'-su6})), 
\bequ
 PP'=\diag(1,-1,-1,-1,1,1),
\eequ 
by which the gauge symmetry is reduced as 
 $SU(6)\rightarrow SU(3)_c\times SU(3)_L\times U(1)$, 
 plays an important role.
The adjoint representation is decomposed as 
\bequ
 {\bf35} \rightarrow 
         ({\bf8} ,{\bf1},+)+({\bf1},{\bf8},+)+({\bf1},{\bf1},+)
        +({\bf3},{\bf{\cc3}},-)+({\bf{\cc3}},{\bf3},-) 
\eequ
in terms of ($SU(3)_c$, $SU(3)_L$, $PP'$). 
Since $PP'=+$ for the Wilson line phases, 
 these d.o.f. are proportional to some of 
 the generators of the reduced symmetry. 
In particular, in this case, the VEV is proportional 
 to one generator of $SU(3)_L$. 
Then, a similar analysis as in section 3 leads the above 
 effective potential.

As for the case of $\eta\eta'=-$, 
 the eigenvalues of 
 $D_y(A_5)^2$ in bulk fields sector 
 become 
\begin{equation}
11\times {(n+1/2)^2\over R^2}, \;\; 
 6\times {n^2\over R^2}, \;\; 
{(n\pm a+1/2)^2\over R}, \;\;\;
 2\times{(n\pm a/2+1/2)^2\over R^2}, \;\; 
 6\times{(n\pm a/2)^2\over R^2}, 
\end{equation}
which lead the effective potential, 
\begin{equation}
 V_\eff^{adj(-)} ={C\over2} \sum_{n=1}^\infty \frac{1}{n^5} \Ll
     6\cos(n\pi a)+2\cos(n\pi (a-1))+
      \cos(n\pi (2a-1)) \Rl.
\end{equation}

For a fundamental representational field with 
 $\eta \eta'=+$, 
 the $U(1)$ charge and $PP'$ are given by 
$$((1/2,+),(0,-),(0,-),(0,-),(0,+),(-1/2,+))^T.$$ 
This means that the 
 eigenvalues are 
\begin{equation}
 {n^2 \over R^2}, \;\;\; 
 3\times{(n+1/2)^2 \over R^2},\;\;\; 
 {(n \pm a/2)^2 \over R^2}, 
\end{equation}
which derive 
 the effective potential, 
\begin{equation}
 V_\eff^{fnd(+)} ={C\over2} \sum_{n=1}^\infty \frac{1}{n^5} \Ll
     \cos(n\pi a)) \Rl.
\end{equation}
On the other hand, 
 a fundamental representational field with 
 $\eta \eta'=-$ has 
$$((1/2,-),(0,+),(0,+), (0,+), (0,-), (-1/2,-))^T,$$
 which means the eigenvalues become 
\begin{equation}
{(n+1/2)^2 \over R^2},\;\;\; 
 3\times{n^2 \over R^2}, \;\;\; 
 {(n \pm a/2+1/2)^2 \over R^2}. 
\end{equation}
Then 
 the effective potential is given by 
\begin{equation}
 V_\eff^{fnd(-)} ={C\over2}\sum_{n=1}^\infty \frac{1}{n^5} \Ll
     \cos(n\pi (a-1)) \Rl. 
\end{equation}
Above calculations reproduce
 the results in Ref.\cite{HHKY}.

\section{The general formula}

Now let us 
 show the general formula of the effective 
 potential in 
 the 5D $SU(N)$ gauge theory 
 with the general boundary conditions. 
In general, the parity operators, $P$ and $P'$, 
 are shown as 
\begin{eqnarray}
P&=&\diag(+,\cdots,+,+,\cdots,+,-,\cdots,-,-,\cdots,-), \nn\\
P'&=&\diag(\underbrace{+,\cdots,+}_{n_+^+},
           \underbrace{-,\cdots,-}_{n_+^-},
           \underbrace{+,\cdots,+}_{n_-^+},
           \underbrace{-,\cdots,-}_{n_-^-}), 
\end{eqnarray}
where $N=n^+_- +n^-_+ +n^+_- +n^-_-$. 
Under the parities, $(P,P')$, 
 the gauge field, $A_\mu$,  
 transforms as 
\begin{equation}
\bordermatrix{
        & n_+^+ & n_+^- & n_-^+ & n_-^- \cr\vsp{0.2}
  n_+^+ & (+,+) & (+,-) & (-,+) & (-,-) \cr\vsp{0.2}
  n_+^- & (+,-) & (+,+) & (-,-) & (-,+) \cr\vsp{0.2}
  n_-^+ & (-,+) & (-,-) & (+,+) & (+,-) \cr\vsp{0.2}
  n_-^- & (-,-) & (-,+) & (+,-) & (+,+) \cr} 
\label{parity-gene}
\end{equation}
which means 
 that $SU(N)$ is broken into $SU(n_+^+) \times SU(n_+^-)$ 
 $\times SU(n_-^+) \times SU(n_-^-) \times U(1)^3$.
Here, $(+,-)$ parts and $(-,+)$ parts have half 
 KK-mode expansion.

The d.o.f. of the Wilson line phases are reside in 
 $(-,-)$ parts in Eq.(\ref{parity-gene}), 
 which are shown as
\begin{equation}
\label{VEVAy}
  \VEV{A_5}=\frac{1}{2gR}\Ls
    \begin{array}{cccc}
      0 & 0 & 0 & \Theta_a \\
      0 & 0 & \Theta_b & 0 \\
      0 & \Theta_b^\dagger & 0 & 0 \\
      \Theta_a^\dagger & 0 & 0 & 0
    \end{array}\Rs.
\end{equation}
The residual gauge freedom reduces 
 the number of the d.o.f. of the Wilson line phases as 
\begin{equation}
 \min(n_+^+, n_-^-) + \min(n_+^-, n_-^+).
\end{equation}
Hereafter, we denote $\min(n_+^+, n_-^-)$ 
 and $\min(n_+^-, n_-^+)$ 
as $A$ and $B$, respectively.
  For example, when $n_+^+ < n_-^-$, $\Theta_a$ can be transformed 
into following form.
\begin{equation}
 \Theta_a=\Ls
   \begin{array}{ccccccc}
     a_1 & 0 & \cdots & 0 & 0 & \cdots & 0 \\
     0 & a_2 & \cdots & 0 & 0 & \cdots & 0 \\
     \vdots & \vdots & \ddots & \vdots & \vdots & \ddots & \vdots \\
     0 & 0 & \cdots & a_A & 0 & \cdots & 0 
   \end{array}\Rs
\end{equation}
Similarly, when $n_+^- < n_-^+$, $\Theta_b$ can be transformed 
into following form.
\begin{equation}
 \Theta_b=\Ls
   \begin{array}{ccccccc}
     b_1 & 0 & \cdots & 0 & 0 & \cdots & 0 \\
     0 & b_2 & \cdots & 0 & 0 & \cdots & 0 \\
     \vdots & \vdots & \ddots & \vdots & \vdots & \ddots & \vdots \\
     0 & 0 & \cdots & b_B & 0 & \cdots & 0 
   \end{array}\Rs
\end{equation}

Let us pick up 
 two non-vanishing VEVs, \eg\ $a_1$ and $a_2$. 
In this case, it is useful 
 to decompose 
 $SU(N)$ into 
 $SU(N-4)\times SU(2)_1 \times SU(2)_2$, 
 where $SU(2)_i$
 is determined by the 
 ``position'' of VEV, $a_i$, 
 in the $SU(N)$ base.
The adjoint representation of $SU(N)$ is decomposed as 
\begin{eqnarray}
  adj. &=& (adj.,{\bf1})+({\bf1},{\bf3})+({\bf1},{\bf1})
          +(fnd.,{\bf2})+(\cc{fnd.},{\bf2}) \nn\\
       &=& [(adj.,{\bf1},{\bf1})+({\bf1},{\bf3},{\bf1})
           +({\bf1},{\bf1},{\bf1})
           +(fnd.,{\bf2},{\bf1})+(\cc{fnd.},{\bf2},{\bf1})] \nn\\
       &&  +[({\bf1},{\bf1},{\bf3})]+[({\bf1},{\bf1},{\bf1})] \nn\\
       &&  +[(fnd.,{\bf1},{\bf2})+({\bf1},{\bf2},{\bf2})]
           +[(\cc{fnd.},{\bf1},{\bf2})+({\bf1},{\bf2},{\bf2})].
\label{adj}
\end{eqnarray}
This suggests 
 the non-vanishing
 eigenvalues of 
 $(T^1_1, T^1_2)$ are 
\begin{eqnarray}
 (\pm1,0),\ (0,\pm1),\ 2\times\Ls \pm\half, \pm\half \Rs,\ 
  (N-4)\times\Ls \pm\half,0 \Rs,\ (N-4)\times\Ls 0,\pm\half \Rs ,
\end{eqnarray}
where $T^1_i$ is the 1st 
 generator $(\tau_1)$ of 
 $SU(2)_i$. 
Furthermore, if these two 
 VEVs reside in the same $\Theta_i$, 
 the components with eigenvalues 
 $(\pm\half, \pm\half)$ exist 
 in the part of $(P,P')=(+,+)$ or 
 $(-,-)$, and therefore 
 have integer KK-expansion.
On the other hand, 
 if these two VEVs reside in different 
 $\Theta_i$, such components exist in the part of 
 $(P,P')=(+,-)$ or $(-,+)$, 
 and therefore have half KK-expansion.
When 
 we deal with more than two non-vanishing VEVs, 
 above 
 observation dealing with two VEVs case 
 is useful.
By taking combinations of 
 above decompositions, 
 we obtain the following 
 general effective potential,%
\footnote{
For $A=0$, $\displaystyle\sum_i^A$ means zero.
} 
\begin{eqnarray}
V_\eff^{adj(+)} 
  &=& {C\over2}\sum_{n=1}^\infty \frac{1}{n^5}
   \Ll \sum_{i,j}^A \cos(n\pi(a_i\pm a_j)) 
     + \sum_{i,j}^B \cos(n\pi(b_i\pm b_j))\RR\nn\\
   &&+ 2\sum_i^A\sum_j^B \cos(n\pi(a_i\pm b_j-1))\nn\\
   &&+ 2\abs{n_+^+-n_-^-}\Ls\sum_i^A\cos(n\pi a_i)
                 + \sum_i^B\cos(n\pi(b_i-1))\Rs\nn\\
   &&\LL+ 2\abs{n_+^--n_-^+}\Ls\sum_i^A\cos(n\pi(a_i-1))
               + \sum_i^B\cos(n\pi b_i)\Rs\Rl,
\label{Vad-g}
\end{eqnarray}
for one d.o.f. of an adjoint 
 representational field. 
As discussed in section 3, 
 the true 
 effective potential 
 is obtained by producting 
 coefficients as, 
 [(fermionic d.o.f.) $-$ 
 (bosonic d.o.f.)] $\times$ 
 Eq.(\ref{Vad-g}).
Especially, the gauge and ghost 
 contributions 
 are obtained by 
 $-3$ $\times$ Eq.(\ref{Vad-g}).

As for the contribution from 
 an adjoint representational field 
 with $\eta \eta'=-$, 
 it is obtained by 
 modifying 
 $Qa_i + Qb_i \rightarrow Qa_i+Qb_i+1/2$, 
 as discussed in section 3. 
The result is 
\begin{eqnarray}
V_\eff^{adj(-)} 
  &=& {C\over2}\sum_{n=1}^\infty \frac{1}{n^5}
   \Ll \sum_{i,j}^A \cos(n\pi(a_i\pm a_j-1)) 
     + \sum_{i,j}^B \cos(n\pi(b_i\pm b_j-1))\RR\nn\\
   &&+ 2\sum_i^A\sum_j^B \cos(n\pi(a_i\pm b_j))\nn\\
   &&+ 2\abs{n_+^+-n_-^-}\Ls\sum_i^A\cos(n\pi(a_i-1))
                          + \sum_i^B\cos(n\pi b_i)\Rs\nn\\
   &&\LL+ 2\abs{n_+^--n_-^+}\Ls\sum_i^A\cos(n\pi a_i)
                             + \sum_i^B\cos(n\pi (b_i-1))\Rs\Rl.
\label{Vad-}
\end{eqnarray}

Next, let us calculate the contribution from 
 a fundamental representational field with $\eta\eta'=+$. 
The parity of the fundamental representation 
 of $SU(N)$, 
 $(P,P')$, is denoted as 
\begin{equation}
\label{VEVAyg}
  \Ls
  \begin{array}{c}
    (+,+) \vsp{0.2}\\
    (+,-) \vsp{0.2}\\
    (-,+) \vsp{0.2}\\
    (-,-) 
  \end{array} \Rs
  \begin{array}{c}
    n_+^+ \vsp{0.2}\\
    n_+^- \vsp{0.2}\\
    n_-^+ \vsp{0.2}\\
    n_-^- 
  \end{array}\; .
\end{equation}
The fundamental representation is decomposed as 
\begin{eqnarray}
  fnd. &=& (fnd.,{\bf1})+({\bf1},{\bf2}), \nn\\
       &=& [(fnd.,{\bf1},{\bf1})+({\bf1},{\bf2},{\bf1})] 
         + ({\bf1},{\bf1},{\bf2}), 
\end{eqnarray}
under the representations of 
 $SU(N-4) \times SU(2)_1 \times SU(2)_2$.
Taking account how 
 the VEV $\VEV{A_5}$ 
 acts on the fundamental 
 representation, Eq.(\ref{VEVAyg}), 
 we can calculate the contribution from 
 a fundamental representational field 
 with $\eta \eta'=+$ as
\begin{eqnarray}
V_\eff^{fnd(+)} ={C\over 2} 
         \sum_{n=1}^\infty \frac{1}{n^5}
     \Ll \sum_i^A \cos(n\pi a_i) + 
          \sum_i^B \cos(n\pi(b_i-1)) \Rl  ,
\end{eqnarray}
and 
 with $\eta \eta'=-$ as
\begin{eqnarray}
V_\eff^{fnd(-)} ={C\over 2} 
          \sum_{n=1}^\infty \frac{1}{n^5}
     \Ll \sum_i^A \cos(n\pi (a_i-1)) + 
             \sum_i^B \cos(n\pi b_i) \Rl .
\end{eqnarray}

For the (anti-)symmetric tensorial 
 representation of $SU(N)$, 
 they are 
 decomposed as
\begin{eqnarray}
  \antisymm{2.5} &=& 
           (\antisymm{2.5},{\bf1})
           +(fnd.,{\bf2})+({\bf1},{\bf1}) \nn\\
                 &=& 
           [(\antisymm{2.5},{\bf1},{\bf1})
     +(fnd.,{\bf2},{\bf1})+({\bf1},{\bf1},{\bf1})]\nn\\ 
   &&+ [(fnd.,{\bf1},{\bf2})+({\bf1},{\bf2},{\bf2})]
            +({\bf1},{\bf1},{\bf1}), \\
  \symm &=& 
   (\symm,{\bf1})+(fnd.,{\bf2})+({\bf1},{\bf3}) \nn\\
        &=& [(\symm,{\bf1},{\bf1})+(fnd.,{\bf2},{\bf1})
            +({\bf1},{\bf3},{\bf1})]\nn\\
        &&+ [(fnd.,{\bf1},{\bf2})+
           ({\bf1},{\bf2},{\bf2})]
           +({\bf1},{\bf1},{\bf3}) ,
\end{eqnarray}
in the base of 
 $SU(N-4)\times SU(2)_1 \times SU(2)_2$. 
The half KK-modes are distributed in a similar way as 
 in 
 the adjoint representation case. 
The effective potential of the anti-symmetric 
 tensor field with $\eta\eta'=\pm$ becomes%
\footnote{
For $A=0,1$, $\displaystyle\sum_{i>j}^A$ represents 
 zero.
} 
\begin{eqnarray}
V_\eff^{\tiny\antisymm{1}(+)} 
  &=& {C\over 2} \sum_{n=1}^\infty \frac{1}{n^5}
   \Ll \sum_{i>j}^A \cos(n\pi(a_i\pm a_j)) 
     + \sum_{i>j}^B \cos(n\pi(b_i\pm b_j))\RR\nn\\
   &&+ \sum_i^A\sum_j^B \cos(n\pi(a_i\pm b_j-1)) \nn\\
   &&+ \abs{n_+^+-n_-^-}\Ls\sum_i^A\cos(n\pi a_i)
                  +\sum_i^B\cos(n\pi(b_i-1))\Rs\nn\\
   &&\LL+ \abs{n_+^--n_-^+}\Ls\sum_i^A\cos(n\pi(a_i-1))
                  +\sum_i^B\cos(n\pi b_i)\Rs\Rl,
\end{eqnarray}
and
\begin{eqnarray}
V_\eff^{\tiny\antisymm{1}(-)} 
  &=& {C\over 2} \sum_{n=1}^\infty \frac{1}{n^5}
   \Ll \sum_{i>j}^A \cos(n\pi(a_i\pm a_j-1)) 
     + \sum_{i>j}^B \cos(n\pi(b_i\pm b_j-1))\RR\nn\\
   &&+ \sum_i^A\sum_j^B \cos(n\pi(a_i\pm b_j)) \nn\\
   &&+ \abs{n_+^+-n_-^-}\Ls\sum_i^A\cos(n\pi (a_i-1))
                   +\sum_i^B\cos(n\pi b_i)\Rs\nn\\
   &&\LL+ \abs{n_+^--n_-^+}\Ls\sum_i^A\cos(n\pi a_i)
                  +\sum_i^B\cos(n\pi (b_i-1))\Rs\Rl.
\end{eqnarray}
On the other hand, 
 the effective potential of the symmetric 
 tensor with $\eta\eta'=\pm$ becomes 
\begin{eqnarray}
V_\eff^{\tiny\symm (+)} 
  &=& {C\over 2} \sum_{n=1}^\infty \frac{1}{n^5}
   \Ll \sum_i^A \cos(2n\pi a_i) + 
                        \sum_i^B \cos(2n\pi b_i) \RR\nn\\
   &&+ \sum_{i>j}^A \cos(n\pi(a_i\pm a_j)) 
     + \sum_{i>j}^B \cos(n\pi(b_i\pm b_j))
     + \sum_i^A\sum_j^B \cos(n\pi(a_i\pm b_j-1)) \nn\\
   &&+ \abs{n_+^+-n_-^-}\Ls\sum_i^A\cos(n\pi a_i)
                +\sum_i^B\cos(n\pi(b_i-1))\Rs\nn\\
   &&\LL+ \abs{n_+^--n_-^+}\Ls\sum_i^A\cos(n\pi(a_i-1))
                      +\sum_i^B\cos(n\pi b_i)\Rs\Rl,
\end{eqnarray}
and 
\begin{eqnarray}
V_\eff^{\tiny\symm (-)} 
  &=& {C\over 2} \sum_{n=1}^\infty \frac{1}{n^5}
   \Ll \sum_i^A \cos(2n\pi (a_i-\half)) 
     + \sum_i^B \cos(2n\pi (b_i-\half)) \RR\nn\\
   &&+ \sum_{i>j}^A \cos(n\pi(a_i\pm a_j-1)) 
     + \sum_{i>j}^B \cos(n\pi(b_i\pm b_j-1)) \nn\\
   &&+ \sum_i^A\sum_j^B \cos(n\pi(a_i\pm b_j)) \nn\\
   &&+ \abs{n_+^+-n_-^-}\Ls\sum_i^A\cos(n\pi (a_i-1))
                    +\sum_i^B\cos(n\pi b_i)\Rs\nn\\
   &&\LL+ \abs{n_+^--n_-^+}\Ls\sum_i^A\cos(n\pi a_i)
          +\sum_i^B\cos(n\pi (b_i-1))\Rs\Rl.
\end{eqnarray}

As for the calculation of 
 SUSY version, 
 the effective potential can be obtained 
 straightforwardly 
 as discussed in section 3. 
The gauge sector contributions 
 are obtained by adding the 
 coefficient, $-4$, 
 and 
 the factor, $(1-\cos(2\pi n\beta))$,  
 in the r.h.s. summation 
 of Eq.(\ref{Vad-g}). 
Here, the factor $-4$ denotes the number of the (bosonic) 
 d.o.f. of massless modes and $(1-\cos(2\pi n\beta))$ 
 is due to the {\it massive} gaugino contributions 
 possessing ${\mathcal O}(\beta /R)$ 
 SUSY breaking masses. 
As for the bulk fields contributions, 
 the effective potential 
 of the SUSY version 
 is obtained by 
 adding the factor $(1-\cos(2\pi n\beta ))$ 
 in the summation $n$ in 
 the non-SUSY effective potential induced from 
 the Dirac fermion of a representation.

For the contributions from 
 higher representations, 
 we can calculate in the same 
 way, since 
 our method only uses 
 the group theoretical analysis.

\section{Summary and discussion}

We show the general formula of 
 the one loop effective potential 
 of the 5D $SU(N)$ gauge theory 
 compactified on an orbifold, $S^1/Z_2$. 
The formula shows 
 the case when 
 there are 
 fundamental, (anti-)symmetric tensor,
 adjoint representational bulk fields. 
Our calculation method is also applicable 
 when there are higher 
 representational bulk fields.
The SUSY version of 
 the effective potential 
 with SS breaking
 can be obtained straightforwardly.  
We have also shown some examples of 
 effective potentials 
 in cases of 
 one VEV with $P=P'$ in $SU(3)$, 
 two VEVs with $P=P'$ in $SU(5)$, 
 three VEVs with $P=P'$ in $SU(6)$ and 
 one VEV with $P\neq P'$ in $SU(6)$. 
All of which are 
 reproduced by our general formula.

We emphasize our method can be also applied 
 to models with a gauge symmetry other than 
 $SU(N)$, such as $SO(10)$ or $E_6$.
It has been difficult to analyze the vacuum 
 structure in those models, because of the hard 
 task of calculating the complicated commutation 
 relations. 
However, our calculation method make it possible to analyze 
 the vacuum structure even in those models. 
We expect novel models can be built through
 such researches.
We will investigate this possibility in another 
 paper.

\vskip 1cm

\acknowledgments

T.Y. is supported by a Grant-in-Aid for 
 the 21st Century COEhCenter for Diversity 
 and Universality in Physicsh. 
This work was supported in part by  Scientific Grants from 
 the Ministry of Education and Science, 
 Grant No.\ 14039207, 
 Grant No.\ 14046208, 
 Grant No.\ 14740164 (N.H.).


\vspace{1cm}

\def\jnl#1#2#3#4{{#1}{\bf #2} (#4) #3}

\def\Zphys{{\em Z.\ Phys.} }
\def\jssc{{\em J.\ Solid State Chem.\ }}
\def\jpsJ{{\em J.\ Phys.\ Soc.\ Japan }}
\def\ptps{{\em Prog.\ Theoret.\ Phys.\ Suppl.\ }}
\def\PTP{{\em Prog.\ Theoret.\ Phys.\  }}

\def\JMP{{\em J. Math.\ Phys.} }
\def\NPB{{\em Nucl.\ Phys.} B}
\def\NP{{\em Nucl.\ Phys.} }
\def\PLB{{\em Phys.\ Lett.} B}
\def\PL{{\em Phys.\ Lett.} }
\def\PRL{\em Phys.\ Rev.\ Lett. }
\def\PRB{{\em Phys.\ Rev.} B}
\def\PRD{{\em Phys.\ Rev.} D}
\def\PRe{{\em Phys.\ Rep.} }
\def\AP{{\em Ann.\ Phys.\ (N.Y.)} }
\def\RMP{{\
em Rev.\ Mod.\ Phys.} }
\def\ZPC{{\em Z.\ Phys.} C}
\def\SCI{\em Science}
\def\CMP{\em Comm.\ Math.\ Phys. }
\def\MPLA{{\em Mod.\ Phys.\ Lett.} A}
\def\IJMPA{{\em Int.\ J.\ Mod.\ Phys.} A}
\def\IJMPB{{\em Int.\ J.\ Mod.\ Phys.} B}
\def\EPJC{{\em Eur.\ Phys.\ J.} C}
\def\PR{{\em Phys.\ Rev.} }
\def\JHEP{{\em JHEP} }
\def\cmp{{\em Com.\ Math.\ Phys.}}
\def\JPA{{\em J.\  Phys.} A}
\def\CQG{\em Class.\ Quant.\ Grav. }
\def\ATMP{{\em Adv.\ Theoret.\ Math.\ Phys.} }
\def\ibid{{\em ibid.} }

\leftline{\bf References}

\renewenvironment{thebibliography}[1]
         {\begin{list}{[$\,$\arabic{enumi}$\,$]}  
         {\usecounter{enumi}\setlength{\parsep}{0pt}
          \setlength{\itemsep}{0pt}  \renewcommand{\baselinestretch}{1.2}
          \settowidth
         {\labelwidth}{#1 ~ ~}\sloppy}}{\end{list}}

\end{document}